\documentclass{amsart}
\usepackage{amssymb}

\title[Comparison of strategies for unification]{Comparison of group-theoretical strategies for unification of $3+1$ structures in physics}
\author{Robert Arnott Wilson}
\date{26th July 2020; this version 3rd August 2020}
\address{Queen Mary University of London}
\email{r.a.wilson@qmul.ac.uk}

\begin{document}
\begin{abstract}
One of the most important advances in our understanding of the physical world arose from
the unification of $3$-dimensional space with $1$-dimensional time into a $4$-dimensional spacetime.
Many other physical concepts also arise in similar $3+1$ relationships, and attempts have been made to unify
some of these also. Examples in particle physics include the three intermediate vector bosons of the weak interaction,
and the single photon of electromagnetism. The accepted unification in this case is the Glashow--Weinberg--Salam model
of electro-weak interactions, which forms part of the standard model. Another example is the three colours of
quarks and one of leptons. In this case, the Pati--Salam model attempts the unification, but is not currently part of the
accepted standard model.

I investigate these and other instances of $3+1=4$ in fundamental physics, to see if a comparison between
the successful and unsuccessful unifications can throw some light on why some succeed and others fail.
In particular, I suggest that applying the group-theoretical methods of the more successful unifications to the less successful ones
could potentially break the logjam in theoretical particle physics.
\end{abstract}
\maketitle

\section{Introduction}
\label{intro}
It so happens that many concepts in physics come in packages of $3+1$, such as 
\begin{enumerate}
\item three dimensions of space and one of time;
\item three dimensions of momentum and one of energy;
\item three colours of quarks and one of leptons;
\item three mediators for the weak  
interaction and one for electromagnetism; 
\item three charged fermions and one uncharged fermion in each generation;
\item three generations of the fundamental fermions and one of the basic bosons;
\item three fermions with charge $-1$ (electrons) and one with charge $+1$ (proton).
\end{enumerate}
Some of these are known to be related, such as the first two, which are dual to each other.
Others are not known to be related, and most are generally assumed to be unrelated. 
Indeed, there is no good reason to think that any one occurrence of the equation $3+1=4$ is
necessarily related to any other. 

Nevertheless, the occurrence of so many instances of this equation in fundamental physics
is rather striking, and
 it is, of course, conceivable 
 that some of them may be related in deeper ways than we currently understand.
Such a relationship need not be physical, but could simply be the application of the same mathematics in
different contexts.
Such multiple uses of the same mathematics are quite common in physics, such as the use of the group
$SU(2)$ as both the spin group in non-relativistic quantum mechanics, and as the gauge group of the
weak force \cite{Zee}. 

In Section~\ref{examples} I describe in more detail the seven examples listed above, in the order given. 
In Section~\ref{maths} I compare various different approaches that have been taken to unifying $3+1$ into $4$
in some of the examples, with a particular emphasis on analysing the differences between the more successful
approaches and the less successful approaches. In Section~\ref{gravity} I speculate on possible consequences
of transplanting the successful strategies to other regimes.
\section{Examples}
\label{examples}
\subsection{Space and time}
The first and most obvious example is the three dimensions of space and one of time. The work of Lorentz, Einstein, Minkowski and others
\cite{Einstein,Minkowski} led to a unification of space and time into an indivisible concept of spacetime, necessary for a proper understanding of light. At a group-theoretical level, the equation $3+1=4$ expresses the extension of the point symmetry group $SO(3)$ of Galilean space to the point 
symmetry group $SO(3,1)$ of Minkowski spacetime. A further extension to $SL(4,\mathbb R)$ or $GL(4,\mathbb R)$ is also used in some approaches to general relativity
\cite{GL4Ra,GL4Rb}, though not all \cite{einstein2,relativity,thooft}.

\subsection{Momentum and energy}
A closely related example is the dual of spacetime in the Hamiltonian sense, that is the three dimensions of momentum plus one of energy. These are united by the same Lorentz group $SO(3,1)$ into the concept of $4$-momentum. 
The groups $SO(3)$ and $SO(3,1)$, that act in essentially the same way on both spacetime and $4$-momentum, 
are lifted to double covers $SU(2)$ and $SL(2,\mathbb C)$ in order to incorporate the spin $1/2$ 
features of fermions that are an essential part of quantum mechanics \cite{Woit,Dirac}.

\subsection{Colours of fundamental fermions}
Another $3+1$ package occurs in the classification of elementary particles into `colours' in quantum chromodynamics \cite{QCD}. 
Here the particles divide into quarks ($3$ colours) and leptons ($1$ colour, or absence of colour). In this case, the colour group in the standard model is $SU(3)$, and an attempt was made in the Pati--Salam model \cite{PatiSalam}
 to make the unification $3+1=4$ by extending the group to $SU(4)$. This attempted unification of quark and lepton colours has not however obtained the same general acceptance as has the unification of space-time, due to difficulties in reconciling it with experiment.

\subsection{Electro-weak gauge bosons}
Other occurrences of a $3+1$ structure occur in the theory (and experiment) of electro-weak interactions. The gauge bosons of this theory split into 3 massive intermediate vector bosons, and 1 massless photon. The group in this case is $U(2)$, which splits as a central product of $SU(2)$ 
with the scalar group $U(1)$. Here, however, there is no larger group that mixes the 3 and the 1. Thus the unification of electromagnetism and the weak interaction is not based on group theory in the same way as the previous examples. There is also a breaking of the symmetry of the intermediate vector bosons into one neutral $Z$ boson and two oppositely charged $W$ bosons.

\subsection{Charge}
There is a similar division of the fundamental fermions in the standard model into one 
neutral type, the neutrino, and three charged types, the electron and the up and down quarks. 
Again we have a broken symmetry between the electrons and the quarks. But in this case, the standard model does not contain any associated symmetry group(s).

\subsection{Generations}
One can combine these two occurrences (bosonic and fermionic) 
of $3+1$ into 3 generations of fermions, and one of bosons, and hence obtain a classification of all the fundamental particles of the standard model, except the 8 gluons and the Higgs boson, into $(3+1) \times (3+1)$. The 8 gluons are normally described as 8 of the 9 dimensions of $3\times3$, and it might make sense to include the Higgs boson as the 9th dimension. Combining $3\times3$ with $(3+1)\times(3+1)$ 
is a problem of unification of the strong force with the electro-weak force, and does not appear at this stage to fit in with the 3+1=4 paradigm. 

\subsection{Integer charge}
Ignoring the colours, quarks, gluons and everything else related to the strong force, the fundamental particles of matter in the electro-weak theory are
the simplest particles that have integer charge, and are therefore best taken as 3 generations of electron plus 1 proton. In this instance again, group theory is not normally used to describe the three generations.

\subsection{Taking stock}
To summarise, I have identified the following seven occurrences of a 3+1 structure:
\begin{enumerate}
\item 3+1 spacetime dimensions;
\item 3+1 dimensions of 4-momentum;
\item 3+1 colours;
\item 3+1 electroweak gauge bosons;
\item 3+1 fermion types (per generation);
\item 3+1 generations of fermions+bosons;
\item 3+1 generations of electrons+proton.
\end{enumerate}
The question then is, whether there is any actual relationship between any of these occurrences of 3+1 structure? This is really two questions, firstly whether there is any mathematical relationship and secondly whether such a mathematical relationship arises from a physical relationship. For example, 
as already mentioned,
the mathematical isomorphism between the spin group $SU(2)$ and the gauge group $SU(2)$ of the weak interaction is very useful, but no-one would regard them as being physically the same group.

\section{The mathematics of unification}
\label{maths}
\subsection{Current theories}
\label{current}
In the standard model and its extensions, several different groups are used to describe different examples of 3+1. The triplet symmetry group is either $SO(3)$, or $SU(2)$, or $SU(3)$, and possible extensions to $3+1$ symmetry include $SO(3,1)$, $SL(2,\mathbb C)$, $U(2)$ and $SU(4)$. The first one describes a genuine extension of the symmetries of spacetime, and $4$-momentum, in special relativity. The second describes the corresponding
extension of the spin group in quantum mechanics. 
All are amply justified by experiment. 

The third group extension is used to join the 3 and the 1 together in the Glashow--Weinberg--Salam
electro-weak theory \cite{GWS}, but does not mix them. 
The mixing has to be bolted on separately, using the Weinberg angle, which is not constant, but varies with the energy scale of the experiment. 
The fourth extension of groups does mix the 3 and the 1 in the Pati--Salam model, 
but this model is unfortunately not consistent with experiment, at least in the usual interpretation. 

\subsection{Lessons from special relativity}
\label{SRlessons}
In both these cases, therefore, it is reasonable to ask whether the mixing might be better described by analogy with the successful mixing of space and time, that is by using (case 4) the group $SL(2,\mathbb C)$ in place of $U(2)$, and (case 3) the groups $SO(3)$ and $SO(3,1)$ in place of $SU(3)$ and $SU(4)$. 
The first of these suggestions is mathematically quite straightforward, as is seen in Section~\ref{EWrevisited}, 
but requires some care in the physical interpretation of the various different algebras and representations in order to produce a result consistent with experiment. The second, however, requires abandoning the group $SU(3)$ used in quantum chromodynamics, and replacing it with the smaller group $SO(3)$.  
Mathematically, we have to replace a complex 3-space by a real 3-space, which may lose important information. 
This approach therefore seems unlikely to provide an adequate basis for extending the standard model from three colours to four.

\subsection{Lessons from general relativity}
\label{GRlessons}
We must 
look to an analogy with general relativity
to provide a bigger group to restore the lost information for the strong force. 
This suggests that the unification group in this case should be $SL(4,\mathbb R)$, extending $SL(3,\mathbb R)$.
Thus we have an additional problem of explaining how (and why) to convert the group $SU(3)$ used in the standard model
into the corresponding split real form $SL(3,\mathbb R)$.
In practice, however, 
the required changes to the standard model are likely to be minimal, 
essentially only involving an occasional factor of $i$ inserted into, or removed from, a formula here and there. 

\subsection{The Pati--Salam model revisited}
\label{PatiSalam}
Now if we use $SL(4,\mathbb R)$ for the four colours, then we can obtain a model of colours that is almost indistinguishable from the Pati--Salam model, except that it uses the split real form rather than the compact real form of the group $SU(4)$. One still has to take care of the physical interpretations, in order to avoid the problems that the Pati--Salam model runs into when compared with experiment. 
Somehow one has to explain, or explain away, the extra $7$ gauge bosons that seem to appear in the extension from $8$ degrees of freedom to $15$.
But at least the mathematics should still work. 

\subsection{Electro-weak unification revisited}
\label{EWrevisited}
First, let us ask what are the consequences of using $SL(2, \mathbb C)$ instead of $U(2)$ for electro-weak unification? 
The dimension of the 
group has increased from 3+1 to 3+3 real dimensions or perhaps 3 complex dimensions. The adjoint representation, as a 6-dimensional real representation, has signature $(3,3)$, and has compact part 
corresponding to the weak gauge group $SU(2)$. It also has a null subspace of dimension 3, which presumably corresponds to a photon, including a parameter for its direction of motion. Moreover, one can split the 6-space as a direct sum of two mutually dual null 3-spaces, which could reasonably be interpreted as the two opposite helicities of photon. At least, this is one possible interpretation. There may be others. 

\subsection{The Weinberg angle}
\label{Weinberg}
Although the group $SL(2,\mathbb C)$ does not contain the group $U(2)$, it does contain copies of $U(1)$ and $SU(2)$
that are disjoint.  
These arise from disjoint Lie subalgebras $u(1)$ and $su(2)$, which can be obtained by first choosing the compact
subalgebra $su(2)$, then a copy of $u(1)$ within it, and finally multiplying (any generator of) this $u(1)$ by a complex number $e^{i\theta}$,
for a suitable angle $\theta$ strictly less than $45^\circ$. This restriction on $\theta$ ensures that the Killing form
of the resulting subalgebra is negative definite, so that the algebra is the algebra of a compact group. 
The angle $\theta$ can then be tuned to agree with the Weinberg angle.

\section{The physics of unification}
\label{gravity}
\subsection{A potential electro-weak-strong unification}
\label{EWS}
Extending to $SL(4,\mathbb R)$ adds another 9 dimensions to the adjoint representation of $SL(2,\mathbb C)$, 
which would normally be interpreted as gauge bosons. Adding new bosons that are not attested experimentally would defeat the whole purpose of the exercise, so we are more or less forced to include 8 gluons and the Higgs boson at this point. 
The subalgebra $sl(3,\mathbb R)$ that corresponds to the gluons is disjoint from the subalgebra $sl(2,\mathbb C)$ that corresponds to the photon and intermediate vector bosons, so that this 
proposed identification of the gauge bosons is not completely impossible. 
In the Pati--Salam model,
 this identification corresponds to identifying the unexplained extra $7$ degrees of freedom as the Higgs boson, the three intermediate vector bosons,
 and $3$ degrees of freedom for the photon, instead of using a separate $SU(2)\times SU(2)$ for these.

\subsection{Mass}
The signature of the adjoint representation is $(9,6)$, so that we can find a subspace of dimension $12$ spanned by two null $6$-spaces, and therefore plausibly consisting of massless particles such as photons and gluons. 
But there is no consistent way to choose all 8 of the gluons to be massless in this scenario. 
The best that can be done is to take the six coloured gluons to be massless. Then the two colourless gluons are forced to be
massive, Higgs-like bosons.
Thus the model is different from the massless model of quantum chromodynamics, but instead it gives us three mass/charge parameters which can be used in various ways depending on how we choose our basis for the adjoint representation.

In this way we obtain a decomposition of the algebra as $H+E+C$, where $H$ is a real $3$-space for the Higgs sector,
$E$ is a complex $3$-space for the electro-weak sector, and $C$ is another complex $3$-space for the strong (colour) sector. 
The space $E$ is closed under the Lie bracket, and forms a copy of $sl(2,\mathbb C)$. The space $C$ is not closed, and generates
a copy of $sl(3,\mathbb R)$ that contains
$2$ of the $3$ dimensions of $H$. This is presumably the mechanism by which the strong force gives mass to hadrons.

\subsection{The parameters of the standard model}
\label{parameters}
Notice that the gauge group $SU(3)$ of the strong force \emph{acts on} the complex $3$-space $C$, but is physically quite distinct from the 
mathematically closely related algebra $sl(3,\mathbb R)$ that is \emph{generated by} $C$. This is analogous to the distinction made earlier between the gauge group
$U(2)$ of the electro-weak force and the Lie algebra $sl(2,\mathbb C)$.

Besides the Weinberg angle already discussed, there are by the usual count $24$ other unexplained parameters of the
standard model of particle physics. This is not the place for a detailed discussion of all of them, but I can give a rough description of
how $24$ parameters arise from the relationship between particular copies of  the Lie subalgebras $sl(2,\mathbb C)$ and $sl(3,\mathbb R)$
of the Lie algebra $sl(4,\mathbb R)$, or equivalently, the relationships between the subspaces $H$, $E$ and $C$ just described.

There are three angles within $H$ that may give rise to important parameters. There are $3\times 6$ real parameters 
relating $H$ to $C$, but there is a factor of $3$ for colours that cannot be measured, reducing this to $6$ real parameters.
Similarly for the relation between $H$ and $E$, with a factor of $3$ for the fact that the direction of spin cannot be measured.
Finally the relation between $E$ and $C$ is expressed by a total of $9$ complex inner products, which would appear to reduce
to $9$ real parameters once the complex conjugation (particle/anti-particle?) symmetry 
is taken into account. This gives a total of $3+6+6+9=24$ further
parameters, exactly the number that is required in the standard model. 

\subsection{Prospects for a theory of everything?}
\label{TOE}
Since we were forced in Section~\ref{EWrevisited}
to include the momentum 
of the photon in the adjoint representation of $SL(2,\mathbb C)$,
the proposed extension to $SL(4,\mathbb R)$ forces us to include the rotation symmetry group $SO(3)$ of space inside
the 
group $SL(4,\mathbb R)$. We then seem to have no choice but to identify this copy of $SL(4,\mathbb R)$ with the 
spacetime copy of $SL(4,\mathbb R)$ coming from general relativity. Hence, if this works, then it not only unifies the electromagnetic, 
weak and strong forces, and the Higgs boson, as well as explaining the running of the Weinberg angle, and many other
parameters, with the energy scale,
but also potentially includes a quantum gravity whose effects are locally indistinguishable from general relativity.

\subsection{Restoring the symmetry}
While the adjoint representation splits as $6+8+1$ consisting of subalgebras $sl(2,\mathbb C)$, $sl(3,\mathbb R)$ and $gl(1,\mathbb R)$
to mimic the standard model, the above discussion suggests that a splitting $6+6+3$ might have better explanatory power,
particularly when it comes to explaining mass. There is no subgroup of $SL(4,\mathbb R)$ that splits the adjoint representation in this way,
although a splitting as $6+9$ can be obtained from $SO(3,1)$, for example. 

There is, however, another representation of $SL(4,\mathbb R)$, on the rank $2$ tensors, that splits as $6+10$ for the full group,
and further as $(3+3')+(1+3+6)$ for $SL(3,\mathbb R)$. If $V$ denotes the real $4$-space on which $SL(4,\mathbb R)$ naturally acts,
and $V'$ denotes its dual, then the adjoint representation is the trace $0$ part of $V\otimes V'$, while the rank $2$ tensors form $V\otimes V$.
As far as the electro-weak forces and special relativity are concerned, $V$ is self-dual, and therefore there is no distinction between
these two cases. But for the strong force and general relativity the difference is important.

This splitting of the rank $2$ tensors is compatible with putting the photon into $3+3'$ as usual, the intermediate vector bosons into
the second copy of $3$, which can thereby be `mixed' with the `left-handed' part of the photon, the Higgs boson into $1$, and the
six coloured gluons into $6$. We then have a ready-made explanation for the symmetry-breaking between the weak and strong forces
as a symmetry-breaking between space and time.

\subsection{Emergence}
Finally, the $25$ parameters of the standard model are all encoded in the geometry of $sl(4,\mathbb R)$.
The above arguments seem to require this algebra to be not just isomorphic to, but actually equal to,
the Lie algebra of infinitesimal changes to the local shape of spacetime derived from general relativity. 
If this is really true, then it implies that all $25$ of the unexplained parameters of the standard model
can be explained in terms of properties of the gravitational field, and vice versa. Looked at one way round,
this provides a mechanism for calculating the standard model parameters from parameters of the
ambient gravitational field. Looked at the other way round, it provides a mechanism for explaining
general relativity as a macroscopic effect of the physics of the standard model as a whole.

On 
one hand, this can be viewed as the emergence of macroscopic behaviour, including gravity,
from the fundamental quantum behaviour of elementary particles described by the standard model.
On the other, it can be viewed as an explanation for the peculiarities of the standard model in terms
of the symmetry-breaking provided by the complicated motion of the experiment through a complicated gravitational field.
Both viewpoints are likely to be extremely useful. Either way, the upshot is that
general relativity and the standard model of particle physics are not incompatible, as is usually thought,
but 
are  \emph{essentially equivalent}.

\section{Discussion}
\label{conclusion}
\subsection{Summary}
There is, of course, no guarantee that any of the analogies that
I have pointed out in this paper will be of any use for physics.
The separate occurrences of $3+1=4$ could all be unrelated both
physically and mathematically. But there is a very limited number of
groups that can act on $3$-dimensional space, and a
very limited number of ways of extending to a group acting on a
$4$-dimensional space. It is therefore actually quite hard to avoid using the
same groups and subgroups more than once.

\subsection{Available groups}
If one restricts to Lie groups acting on a $3$-space, then among simple groups
only real or complex forms of
$A_1$ and $A_2$ are available, while on a $4$-space one can also
have forms of 
$A_3$ and $B_2$.
If the spaces are real, this restricts us to the groups $SO(3)$, $SL(3,\mathbb R)$,
$SU(2)$, $SL(2,\mathbb C)$,
$SO(3,1)$, 
$Sp(4,\mathbb R)$ and $SL(4,\mathbb R)$.
Most of these groups have been discussed and compared in this paper.
The remaining one is $Sp(4,\mathbb R)$, which is a double cover of $SO(2,3)$. This is
a $10$-dimensional group that
is reminiscent of a real analogue of the complex Georgi--Glashow model \cite{SU5GUT},
built on the embedding of $SU(2)\times SU(3)$ into $SU(5)$. It may well be useful as an
intermediate stage between $SL(2,\mathbb C)$ and $SL(4,\mathbb R)$.

There are also some non-simple groups of type $A_1A_1$, namely
 $SO(4)$, $SO(2,2)$ and $Spin(2,2)$, that are
subgroups of $SL(4,\mathbb R)$, and that are therefore 
likely to be useful in some way.
For example, the group
$SO(4)$ is a product of two copies of $SU(2)$, 
and is likely to be useful in exploring the implications of this work for the foundations of
quantum mechanics. 
The other two cases look like real analogues of the group $SU(2,2)$ that underpins
Penrose's twistor theory \cite{twistors,penrose,twistorlectures}, so may again be useful for relating the two approaches.

\subsection{Further work}
The remarkable success of special and general relativity in unifying the $3+1$ of space and time into
the $4$ of spacetime therefore strongly suggests that trying to use the same underlying group theory
to unify other occurrences of $3+1$ in physics is likely to be a good strategy. Certainly, the group theory (if any)
that has been used in the many other attempts at such unification is different, and this might explain their
relative lack of success. In forthcoming work \cite{remarks,uses}
I consider in much more detail some ways in which the proposed strategy might be
followed, and what it might tell us about some of the more intractable problems in the
foundations of physics.

\subsection{Comparison with other approaches}
The approach taken here has little or nothing in common with any of the `mainstream' 
approaches to unification
of theories of fundamental physics, such as supersymmetry and string theory \cite{string,SUSY} or loop quantum gravity \cite{loopQG}.
Nor is it closely related to any of the many other interesting approaches such as \cite{
ManogueDray,Furey}, apart from sharing 
with these an emphasis on the primacy of algebra and group theory.

My approach is based on the conviction that 
the key issue is to sort out the underlying group theory first. 
Rather than taking the usual historical approach of trying all possible groups until
one is found that works, I take the view that it is better to use the theory of groups,
Lie algebras
and representations to prove that there is only one possible group to use.
This is a deeply unfashionable point of view, but a change in fashion 
may 
 be what the
subject 
needs at this point.

The idea that gravity may have an effect on the structure and properties of elementary particles
goes back to a paper of Einstein published more than $100$ years ago \cite{Einsteinstructure}.
There he describes how the equations of general relativity can tell us about the internal structure
of the electron and the proton, but says there are not enough equations to determine the exact nature
of the charge distribution within the particle. These ideas were forgotten, ignored or dismissed,
and not used in the
development of theories of the strong force $50$ years later. The lack of sufficient equations is what
is today called asymptotic freedom \cite{freedom1}. 
Perhaps Einstein 
was right all along.

\end{document}